\documentclass[aps,prl,twocolumn,showpacs,floats]{revtex4}
\usepackage{epsfig}
\usepackage{amsmath}
\newcommand{\be}{\begin{equation}}
\newcommand{\ee}{\end{equation}}
\newcommand{\bea}{\begin{eqnarray}}
\newcommand{\eea}{\end{eqnarray}}
\newcommand{\ba}{\begin{array}}
\newcommand{\ea}{\end{array}}
\newcommand{\nn}{\nonumber}

\newcommand{\al}{\alpha}
\newcommand{\sig}{\sigma}

\newcommand{\eps}{\epsilon}

\newcommand{\ua}{\uparrow}
\newcommand{\da}{\downarrow}
\newcommand{\ket}[1]{| #1 \rangle}
\newcommand{\bra}[1]{\langle #1 |}
\begin{document}

\title{\bf Detecting Entanglement Using a Double-Quantum-Dot Turnstile}

\author{
M. Blaauboer$^{1}$ and D.P. DiVincenzo$^{1,2}$
}

\affiliation{$^{1}$ Kavli Institute of Nanoscience, Delft University of Technology,
Lorentzweg 1, 2628 CJ Delft, The Netherlands\\
$^{2}$ IBM T.J. Watson Research Center, P.O. Box 218, Yorktown Heights NY 10598, USA
}
\date{\today}

\begin{abstract}
We propose a scheme based on using the singlet ground state of an
electron spin pair in a
double-quantum-dot nanostructure as a suitable setup for
detecting entanglement between electron spins via the measurement
of an optimal entanglement witness. Using time-dependent gate
voltages and magnetic fields the entangled spins are separated and
coherently rotated in the quantum dots and subsequently detected
at spin-polarized quantum point contacts. We analyze the coherent
time evolution of the entangled pair and show that by counting
coincidences in the four exits an entanglement test can be done.
This setup is close to present-day experimental possibilities and
can be used to produce pairs of entangled electrons ``on demand''.

\end{abstract}

\pacs{03.65.Ud, 03.67.Mn, 73.63.Kv}
\maketitle

One of the challenges in present-day condensed-matter nanophysics
is the controlled production of high-fidelity entangled states of
separated electrons for use in quantum information processing.
Such an operation is difficult, largely because of the strong
interactions between electrons, making it difficult to isolate and
coherently manipulate (single) entangled pairs.  But a number of
promising ideas have recently been put forward, involving Cooper
pairs extracted from a superconductor~\cite{leso01,samu03},
two electrons in a single or double quantum
dot~\cite{loss00}, electrons scattered at a
local magnetic impurity~\cite{cost01}, or electrons scattered at
one or more beam splitters~\cite{been03}. Several of these
ideas have led to explicit proposals for detection of entangled
pairs via a test of Bell's inequality by measuring
noise~\cite{samu03,chtc02,been03}.\newline

Here we propose a new scheme for creating entangled electron pairs
using a double quantum dot, rotating their joint state using
electron spin resonance (ESR) manipulations, separating them into
two different quantum channels using a turnstile, and 
collecting them using coincidence detection (i.e. direct counting
of electron pairs) rather than with noise measurements. Our scheme
involves singlet pairs of electrons and controlled manipulation of
their spins, and thereby forms an electronic counterpart of the
photon scheme of Aspect {\it et al.}~\cite{aspe82}. Compared to
the latter, however, our scheme has the advantage that
entanglement can be created in a fully controllable way (``on
demand'').

Furthermore, we propose to characterize the entanglement of the
pairs quantitatively, using the rigorous entanglement tests
recently developed in quantum information theory.  In particular,
we show explicitly how to implement an optimal {\em entanglement
witness}~\cite{TerhalW}, that is, the measurement of a particular
Hermitian operator $W$.  $W$ is designed so that the sign of the
expectation value of the measurement $p={\rm Tr}W\rho$ signals the
presence of entanglement; the more negative $p$ is, the more
entangled the state is.  The measurement is implemented by
separate, local electron spin measurements on the two emitted
electrons, much in the style of a Bell-inequality test. (In fact,
a Bell test is a special case of a particular (non-optimal)
entanglement witness measurement~\cite{TerhalW}.)

\begin{figure}
\centerline{\epsfig{figure=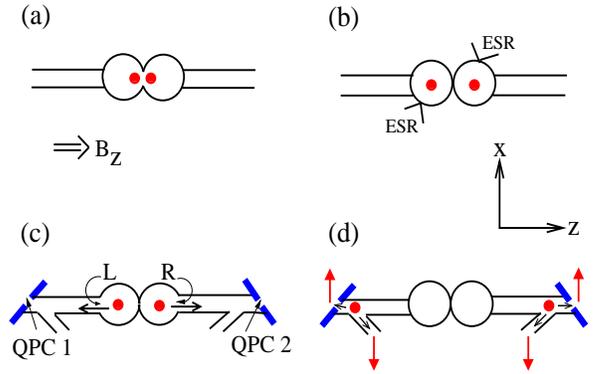,height=5.cm,width=0.9\hsize}}
\caption[]{Schematic top view of the double-quantum-dot setup as
discussed in the text. } \label{fig:system}
\end{figure}

More precisely, the system we will consider consists of two
coupled quantum dots in a parallel magnetic field $B_z \hat{z}$
which are connected to two quantum point contacts (QPCs) via
empty quantum channels, see Fig.~\ref{fig:system}.  The double dot is
initially occupied by two electrons (Fig.~\ref{fig:system}(a)) in
their lowest energy state, the singlet state~\cite{singlet}. The
gate between the two dots is then adiabatically closed, so that
the electrons become separated and one dot is occupied by an
electron with spin-up and the other by one with spin-down. The two spins
do not interact anymore and are independently rotated by ESR
fields (Fig.~\ref{fig:system}(b)).  (The
coherent rotation of spins by ESR fields is the equivalent of the
rotation of the polarizers in optical Bell
experiments~\cite{aspe82}.)  After spin rotation, the electrons
are emitted into the quantum channels by opening gates L and R
(Fig.~\ref{fig:system}(c)) and scattered at quantum point contacts
QPC 1 and QPC 2. In a parallel magnetic field and for conductances
$G_{\rm QPC1 (QPC2)} \leq  e^2/h$ these QPC's are
spin-selective~\cite{poto02}, transmitting electrons with spin-up
and reflecting those with spin-down (Fig.~\ref{fig:system}(d)).
The transmitted and reflected electrons are separately detected in
the four exits.

As Guehne {\em et al.} have recently discussed~\cite{gueh02}, the
optimal entanglement witness operator for detection of a singlet,
$W=\ket{\phi}\bra{\phi}^{PT}$ (where $PT$ denotes the matrix
partial transpose and
$\ket{\phi}=1/\sqrt{2}\,(\ket{\uparrow\uparrow}+\ket{\downarrow\downarrow})$),
can be achieved by having the two electron spins simultaneously
rotated either by $\pi/2$ around the $x$-axis in spin space,
$\pi/2$ around the $y$ axis, or leaving them unrotated.  One
measures the probability of anticoincidences (spins opposite) $A$
for the $y$-rotation cases, and the probability of coincidences
$C$ averaged over the other two settings.  Then, the value of the
entanglement witness measurement ${\rm Tr}W\rho$ is given by
$(2C-A)/2$. While this prescription is well known 
in the quantum information community, it has not been
well appreciated in the electron physics community that these more
precise, quantitative measures of entanglement are no more
difficult than the ``standard" Bell test~\cite{clau69}.

We now analyze the dynamics of the two spins from the
moment they are separated, each occupying one of the two dots,
until both have been detected in one of the four exits. Using a
density matrix approach, this process is represented by the time
evolution of a set of quantum states due to coherent (ESR)
and incoherent (dissipative and tunnel) couplings.
The incoherent processes are incorporated
by phenomenological decay rates and we estimate these rates using
experimentally measured values (for quantum dots) and Fermi's
golden rule (for quantum channels). By solving the resulting set
of equations the probabilities $P_{\sigma \sigma^{\prime}}(t)$ to
simultaneously detect pairs of spins ($\sig$,$\sig^{\prime}$) with
$\sigma, \sigma^{\prime}$$\in$$\{\ua,\da \}$ are
obtained. Our main conclusion is that detection of entanglement
in this double dot turnstile via an entanglement witness measurement,
including Bell tests, is feasible under
realistic circumstances. We first present the model, followed
by a discussion of the results.

In the setup as depicted in Fig.~\ref{fig:system} each electron is
assumed to be either in a dot, in a channel or detected. This leads to a set of 36
possible quantum states represented by a 36$\times$36 density matrix
$\rho(t)$. This set consists of all possible combinations $A\sigma
B\sigma^{\prime}$, with $A$$\in$$\{D,C,X \}$ and $\sigma$$\in$$\{\ua,\da\}$
indicating the position ($D$$=$dot, $C$$=$channel and $X$$=$exit) and spin
direction along $\hat{z}$ of the electron which started out
in the left dot, and $B$$\in$$\{D,C,X \}$ and $\sigma^{\prime}$ representing
the position and spin direction of the electron which started out
in the right dot~\cite{channel}.
The time evolution of the density matrix elements
$\rho_{nm}(t)$ is given by the master equations
\begin{subequations}
\bea
\dot{\rho}_{nn}(t) & = & -\frac{i}{\hbar} \left[ {\cal H}(t), \rho(t)
\right]_{nn} + \nn \\
& & \hspace*{0.3cm} \sum_{m \neq n} \left( W_{nm}\, \rho_{mm}(t) - W_{mn}\,
\rho_{nn}(t) \right) \\
\dot{\rho}_{nm}(t) & = & -\frac{i}{\hbar} \left[ {\cal H}(t), \rho(t)
\right]_{nm} -
V_{nm}\, \rho_{nm}(t) \hspace{0.1cm} n \neq m
\label{eq:Masterb}
\eea
\label{eq:Master}
\end{subequations}
\noindent for $n,m$ $\in$ $\{1,\dots,36\}$. The Hamiltonian ${\cal
H}(t)$ describes the coherent evolution of the system due to the
ESR fields and is given by, for two oscillating magnetic fields
$B_{xL} \cos(\omega t)\, \hat{x}$ and $B_{xR} \cos(\omega t)\,
\hat{x}$ applied to the left and right dots
respectively~\cite{sin}, \bea {\cal H}(t) &= &{\cal H}_{0} -
\frac{1}{2} g^{*} \mu_{B} \cos(\omega t) \! \! \! \! \! \! \!
\sum_{\stackrel{M,N \in \{L,R \}}{M\neq N}} \! \! \! \! \! \!
(B_{xM} + \eps B_{xN}) \bar{\sigma}_{xM}. \label{eq:Hamiltonian}
\eea Here ${\cal H}_{0}$ is a diagonal matrix containing the
energies $E_{n}$ ($n=1,\dots,36$) of each state, $g^{*}$ the
electron g-factor, $\mu_B$ the Bohr magneton and
$\bar{\sigma}_{xL(R)}$ a 36$\times$36 matrix with elements
$(\bar{\sigma}_{xL(R)})_{ij}=1$ for each pair of states $(i,j)$
that are coupled by the oscillating field $ B_{xL(R)}$ and zero
otherwise. For example, for the four states in which both
electrons are in a dot $D$$\ua$$D$$\ua$, $D$$\ua$$D$$\da$,
$D$$\da$$D$$\ua$, $D$$\da$$D$$\da$ (labelled 1, 2, 3, 4) the
energies are given by $E_1 = 2 E_{\ua} + E_C$, $E_2 = E_3 =
E_{\ua} + E_{\da} + E_C$ and $E_4 = 2 E_{\da} + E_C$ in terms of
the single-particle energies $E_{\ua}$ and $E_{\da}$ and the
charging energy $E_C=e^2/C$, and the matrix elements
$(\bar{\sigma}_{xL})_{ij} = 1$ for pairs (1,3), (3,1), (2,4),
(4,2) and zero otherwise. The parameter $\eps$, with $0 \leq
\eps<1$, represents the relative reduction of the field which is
applied to one dot at the position of the spin in the other dot.
This is discussed in more detail towards the end of this paper.
\newline

Turning to the transition rates $W_{nm}$ (from state $m$ to $n$)
in Eqs.~(\ref{eq:Master}), we distinguish between two kinds of
transitions: (1) spin-flip transitions between two states that
differ by the direction of one spin only and (2) tunneling
(without spin-flip) between states that involve adjacent parts of
the system, i.e. from dot to channel and from channel to exit. The
latter are externally controlled by opening and closing gates. The
former are modeled by the phenomenological rate $1/T_{1,\al}
\equiv W_{\al \ua \da} + W_{\al \da \ua}$ with $\al \in \{D,C \}$
for spin flips in a dot or channel. Here the $W$s depend on the
Zeeman energy $|g^{*}| \mu_B B_z$ and temperature $T$ via detailed
balance $W_{\al \ua \da}/W_{\al \da \ua}= e^{|g^{*}| \mu_B
B_z/k_{B} T}$, so that \be W_{\al \ua \da\, (\da \ua)} =
\frac{1}{T_{1,\al}}\, \frac{1}{1 + e^{- (+)|g^{*}| \mu_B B_z
/k_{B} T}} \hspace*{0.3cm} \al \in \{D,C \}. \ee

Finally, the spin decoherence rates $V_{nm}$ in Eqs.~(\ref{eq:Masterb})
for states $n$ and $m$ such that either $n$ or $m$ or both correspond to
a state in which at least one electron is located in a channel and none in the exits, are given by
\bea
V_{nm} =
\frac{1}{T_{2,C}} + \frac{1}{2} \sum_{j\neq n,m} ( W_{jn} +
W_{jm}).
\label{eq:decoherence}
\eea
The coherence between state $n$ and $m$ thus not only depends on
the intrinsic spin decoherence time $T_{2,C}$, but is also reduced by 
tunneling processes from dot to channel and from channel to exit~\cite{enge02}.

With the above ingredients, the coupled equations
(\ref{eq:Master}) can be solved analytically. Details of this
calculation are presented elsewhere~\cite{blaa04}. The solution
obtained is exact under 3 assumptions: 1) the time evolution
during ESR in the dots is decoupled from the time evolution in the
channels and exits. Physically, this means that no 
tunneling occurs out of the dots during the ESR rotations. 2) Once the
electrons are in a channel they cannot tunnel back into the dots,
i.e. back reflection is neglected. 3) Once the electrons are
in one of the exits they are immediately absorbed. With these
realistic assumptions, the master equations are solved in 3 steps:
the time evolution during ESR applied to the left and right dots 
and the time evolution after the gates to the
quantum channels have been opened. During each step only part of
the quantum states are evolving in time, while the others 
remain unchanged. This simplifies the procedure to obtain an
analytical solution.

In order to characterize the entanglement of the two electron
spins, we are interested in the time evolution of the
probabilities that both spins have been detected. Implementing the
optimal entanglement witness then consists of measuring $C(t)$
($A(t)$), the cumulative probabilities that a coincidence
(anticoincidence) count has happened at time $t$, for the spin
rotations as discussed above. These probabilities depend on the
angles of rotation during the two ESR processes, on the tunneling
rates $W$, and on the $T_1$ and $T_2$ times in the dots and
channels. The full expressions are given in Ref~\cite{blaa04}.

\begin{figure}
\centerline{\epsfig{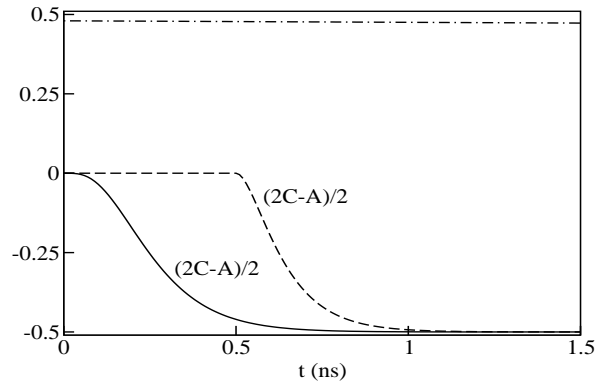}}
\caption[]{The entanglement witness probability $(2C-A)/2$ as a function of time for travel times across the channels $t_{\rm travel}=0.5$ ps (solid curve) and $t_{\rm travel}=0.5$ ns (dashed curve), and the absolute value of the off-diagonal matrix element $|\rho_{\rm off-diagonal}(t)|$ (dot-dashed curve), see the discussion in the text. $t_{\rm travel}$ is assumed to be the same in both channels. Parameters used are $W_T=10^{10} s^{-1}$ for the tunnel rate out of the dots, $W_E=10^{10} s^{-1}$ for the tunnel rate into the exits, and $V_C=10^7 s^{-1}$ (corresponding to $T_{2,C}=0.1\mu$s) for the spin decoherence rate in the channels.
}
\label{fig:C}
\end{figure}

Fig.~\ref{fig:C} shows the probability $(2C-A)/2$ for the spin
rotations discussed before as a function of time $t$, where $t=0$
corresponds to the moment at which the gates to the channels are
opened. It also shows the (very slight) decay of the off-diagonal
matrix element $|\rho_{\rm off-diagonal}(t)|$ which describes the
coherence of the entangled singlet pair in the isolated double
dot, i.e. in the absence of tunnel coupling to the channels
($|\rho_{\rm off-diagonal}(t)|=\frac{1}{2}$ for the fully coherent
pair). By comparing this matrix element with the entanglement
condition $(2C-A)/2 < 0$, the critical amount of decoherence above
which the experiment does not work anymore can be estimated. For
the (realistic) parameters used in Fig.~\ref{fig:C} the
decoherence at the time when the probability that both spins have
been detected has approached 100\% (at $t\approx 1$ ns) is
sufficiently small for detection of entanglement to be possible.

Finally, we briefly discuss the estimation of the $T_1$ and $T_2$ times. In order for the entanglement detection to be successful, spin coherence of each detected pair must be preserved from the moment of creation of the singlet in the double dot until the two electrons have been detected. It is thus essential to investigate the time scales $T_2$ and $T_1$ for decoherence and dissipative processes in the dot and channels. Starting with the dot, $T_{1,D}$ for a single spin has recently been experimentally measured and  found to be $T_{1,D} = 0.85$ ms at $B=8 T$~\cite{elze04}. The spin decoherence time $T_{2,D}$ has not been measured yet for a single spin in a quantum dot but is expected to lie between 100 ns $\ll T_{2,D} < 10\, \mu$s~\cite{kikk99,khae03}. In the quantum channels we have estimated a lower bound for $T_{1,C}$ by calculating the inverse relaxation rate due to emission or absorption of a phonon (in the absence of a spin-flip~\cite{relaxation}) using Fermi's golden rule. We obtain $T_{1,C} \geq \frac{2.3\, \cdot\, 10^{-19}}{L[m]}$ s for B=8T and $T_l$ = 100 mK, with $L$ the channel length.

In order to estimate $T_{2,C}$ we apply the calculations in Ref.~\cite{khae03} for the decoherence time of a single spin in a quantum dot due to hyperfine interaction~\cite{shen04} to a quantum channel. This results, just as for a dot, in the bounds $\hbar \sqrt{N}/A < T_{2,C} <  \hbar N/A$, with $N$ the number of nuclei in the channel and $A$ the mean hyperfine
interaction strength. Using $N \sim 10^{5}$ in a channel of length 1$\mu$m and $A=90 \mu$eV for GaAs we find $T_{2,C} > 5$ ns.

We see that in the dots the decoherence time is the limiting time
scale: separating the entangled pair and rotation by ESR fields
has to happen within $T_{2,D}$. The time it takes to adiabatically
close the barrier between the two dots $T_{\rm close}$ can be
obtained from the requirement $|\al_i^{\rm max}/\omega_i^{\rm
min}|^2 \ll 1$, where $\al_i^{\rm max}$ is the maximum angular
velocity of the dot eigenstate $\psi_i(t)$ during the closing and
$\omega_i$ the minimum Bohr frequency~\cite{mess58}. Modeling the
quantum dots with a parabolic confining potential $V(t) =
\frac{1}{2} m^{*} \omega_D^2(t) r^2$, we estimate $T_{\rm close} >
\frac{\hbar \omega_D^2(0)}{|g^{*}| \mu_B B_z
(2\omega_D^2(0)+\omega_c^2)}$ $\sim \hbar / E_{ST} \sim 10\, {\rm
ps} \ll T_{2,D}$ with $E_{ST}$ the singlet-triplet energy
splitting~\cite{burk99}. Assuming ESR fields of 1 mT, the rotation
time $T_{\rm rot} \sim 30$ ns (for rotations by $\pi/2$), so
provided $T_{2,D} \gg$ 100 ns both separation of the pair and two
rotations can be performed within the decoherence time $T_{2,D}$.
For quantum channels that are 0.1 $\mu$m long the travel time of
an electron, assuming ballistic transport, is
$t_{\rm travel} \sim L/v_{ch} \sim 0.5$ ps, which is less than
both $T_{1,C}$ and $T_{2,C}$. The arrival of consecutive spin
pairs at the QPC's is then separated by $\sim 10^{-7}$\, s, which
corresponds to a current signal $\sim 1-10$\, ps. This is
observable and forms a more transparent way of detection than
noise measurements. By measuring the current in the four exits in
real time~\cite{vand04} for repeated runs of the experiment, the
coincidence probabilities $A$ and $C$ can be determined.

Let us finish with some remarks on the detectors and the ESR rotations. It has recently been demonstrated that a QPC which is set to transmit only a single mode becomes spin-polarized in a parallel magnetic field~\cite{poto02}: it only transmits electrons with spin-up and reflects those with spin-down, which is due to the fact that the spin-down electrons in the Fermi sea see a higher barrier at the QPC than the spin-up electrons. This is also true for a single electron that arrives at the QPC in an otherwise empty channel, provided that the electron has not relaxed to the bottom of its energy band. In the latter case, the QPC barrier height for a spin-up and a spin-down electron is the same. As long as the time the electron spends in the channel is sufficiently less than $T_{1,C}$, the QPC detectors are able to distinguish between the two spin directions~\cite{detectors}.

Finally, turning to the spin rotations, currently-available ESR fields cannot be applied to one of two adjacent dots only, as represented by the parameter $\eps$ in Eq.~(\ref{eq:Hamiltonian}). It is therefore not possible to use each field to coherently rotate only {\it one} spin, as was done with polarizers in the analogous photon experiment~\cite{aspe82}. It is, however, nevertheless possible - by using both ESR fields during each of the two rotation-intervals in the absence of exchange coupling between the two spins - to rotate both spins independently. Alternatively, one could use two well-separated additional dots, each of which is located between the double dot and a detector, to perform independent ESR rotations.

In conclusion, we have presented a scheme for detecting
entanglement of electron spins in a double-quantum-dot turnstile.
Using a carefully timed series of pulsed gate and ESR operations,
pairs of entangled electron spins are separated and coherently
manipulated in the quantum dots and subsequently detected at two
spin-polarized QPC's. We have analyzed the time scales for
spin decoherence $T_2$ and relaxation $T_1$ and
demonstrate that under realistic experimental conditions the
evolution from creation to detection of each pair can be completed
within the coherence time, thus allowing for a quantitative
entanglement test
to be performed. Moreover, due to the high level of external
control over the dynamics, this setup can be used to produce
pairs of entangled electrons on demand.

Stimulating discussions with L.P. Kouwenhoven, L.M.K. Vandersypen and B.M. Terhal
are gratefully acknowledged. This work has been supported by the Stichting voor
Fundamenteel Onderzoek der Materie (FOM), by the Netherlands Organisation for
Scientific Research (NWO) and by the EU's Human Potential Research Network under
contract No. HPRN-CT-2002-00309 (``QUACS''). DPDV is supported in part by the
NSA and ARDA through ARO contract numbers W911NF-04-C-0098 and DAAD19-01-C-0056.


\begin{thebibliography}{99}
\bibitem{sin}
For rotations around $\hat{y}$, oscillating magnetic fields
$B_{yL}\sin(\omega t) \hat{y}$ and $B_{yR}\sin(\omega t) \hat{y}$
are used.
\bibitem{detectors}
Alternatively, additional dots could be used as spin detectors.
\end{thebibliography}
\end{document}